\documentstyle[12pt]{article}
\newcommand{\beq}{\begin{eqnarray}}
\newcommand{\eeq}{\end{eqnarray}}
\newcommand{\D}{{\Delta}}

\newcommand{\no}{{\nonumber}} 
\newcommand{\f}{\frac}

\begin{document}
\topmargin 0pt
\oddsidemargin -3.5mm
\headheight 0pt
\topskip 0mm \addtolength{\baselineskip}{0.20\baselineskip}
\begin{flushright}
 CCNY-HEP-01/11 \\
   hep-th/0201002
\end{flushright}
\vspace{0.3cm}
\begin{center}
  {\large \bf  Non-Commutative Space-Times, Black Hole, and Elementary Particle}
\end{center}
\vspace{0.3cm}
\begin{center}
 Mu-In Park\footnote{Electronic address:
muinpark@yahoo.com}
\\{ Department of Physics, City College of the CUNY}\\{ New York, NY 10031}
\end{center}
\vspace{0.3cm}
\begin{center}
  {\bf ABSTRACT}                                                               
\end{center}
It is shown that {\it elementary black hole} can not be distinguished
from an elementary particle in the non-commutative
space-times (space/space and space/time) at the Planck scale.
But, the non-commutative space-times can not be ``directly'' measured
in the elementary black hole system. 
A $time-varying$ non-commutative parameter $\theta(t)$ is suggested in
accordance with the time-varying-$G$ scenario. 
By identifying the elementary black hole with an elementary particle, the large hierarchy between the weak scale and Planck scale is
naturally understood. 
For {\it large black hole}, the discrete spectrum of the horizon
area is derived by identifying the black hole horizon with a non-commutative sphere.
\vspace{0.1cm}
\begin{flushleft}
Keywords: Non-Commutative Geometry, Black Holes.\\
7 Jan 2002 \\
\end{flushleft}
\newpage
Recently, there have been notable studies on the non-commutative
space-times in diverse aspects \cite{Conn}. However, it seems still
unclear whether this is just a hypothetical extension of the
usual theory as the commutative space-times are generalized to the
non-commutative space-times inspired by the string theory, or a new
physical paradigm which governs the space-time quantum.
Moreover, we don't even know how small the non-commutative parameter is. 

In this paper, I show that how the non-commutative $space-times$\footnote{It has
  been well-known that the lowest Landau level has the non-commutative
  spatial coordinates \cite{Dunn}. More recently, it has been proposed
  that the Chern-Simons theory on the non-commutative $space$ is an
  {\it effective} description of the fractional quantum Hall effect
  \cite{Suss:01,Nair:01}.} (space/space and space/time) can provide
  new insights on our physical world which is sometimes mysterious: If one identify the non-commutative space/space, space/time
  parameter $\sqrt{\theta}$ as the Planck 
length, the elementary black hole, which is smaller than the Planck scale, can not be distinguished from the
elementary particle such as the elementary black hole may be
identified as an elementary particle. But, it is found that the
  non-commutative space-times can not be ``directly'' measured in the
  elementary black hole system. A $time-varying$ non-commutative parameter $\theta(t)$ is suggested in
accordance with the time-varying $G$ scenario.  
From the identification of the elementary black hole with the elementary
  particle, the large hierarchy between the weak scale and Planck
  scale is naturally understood. A plausible way of resolution for the
  instability problem is proposed within the non-commutative quantum
  mechanics. For the large black hole, which is larger than
  the Planck scale, the discrete spectrum of the horizon
area may be naturally understood from a non-commutative sphere.

{\it Elementary black hole}:
Let me start by considering an elementary black hole whose horizon
radius $r_+$ is smaller than the Planck length $l_{\hbox{\scriptsize
    Planck}}$
\beq
\label{lp}
l_{\hbox{\scriptsize Planck}}=\sqrt{\f{G \hbar}{c^3}}\sim 10^{-33}~ \mbox{cm}.
\eeq  

But since the black hole (with mass $m$) has a horizon, according to the general relativity, at
\beq
\label{r+=ls}
r_+ \sim l_{\hbox{\scriptsize Schwarzschild}},
\eeq
where $l_{\hbox{\scriptsize Schwarzschild}}$ is the radius of the
Schwarzschild-black-hole horizon:
\beq
\label{ls}
l_{\hbox{\scriptsize Schwarzschild}}=\f{2 Gm }{c^2},
\eeq
the condition of the elementary black hole becomes
\label{ls<lp}
\beq
l_{\hbox{\scriptsize Schwarzschild}} \leq l_{\hbox{\scriptsize Planck}}   
\eeq
and then one finds a condition for the mass for the elementary black
hole 
\beq
\label{m<mc}
m \leq m_{NC}
\eeq
with a critical mass
\beq
\label{mc}
m_{NC} =\f{1}{2} \sqrt{\f{c \hbar}{G}} \sim 10^{-5}~\mbox{g} ,
\eeq
which is (half) of the Planck mass.

Now, let me consider the elementary black hole on a two dimensional
non-commutative quantum plane with the coordinates $x_1,x_2$:
\beq
\label{xx}
[x_1, x_2 ] &=&i \theta.
\eeq
The non-commutativity parameter $\sqrt{\theta}$ defines a length scale
in which the non-commutativity is relevant. On the other hand, it is well-known
in the black hole physics that the Planck
length is a natural length below which the
usual continuous space and time may be broken down
\cite{Beke:73,tHoo:96}. I take this as a fundamental assumption in
this paper. So, let me take $\sqrt{\theta}= l_{\hbox{\scriptsize Planck}}$ such as the commutation
relation (\ref{xx}) implies that
\beq
\label{dxdx}
\Delta x_1 \Delta x_2 \geq l_{\hbox{\scriptsize Planck}}^2 /2.
\eeq
But there is no restriction on the accuracy of the measurement for one
coordinate at $one$ time so far.

Then, let me consider a measurement process for the coordinates of the
elementary black hole on the quantum plane. To this end, one first
note that the genuine size of the black hole is 
the horizon size $r_+$ since the black hole horizon is a boundary of our
observable world. So, if the elementary black hole had a best localized position $before$ the
measurement, then the position will be given $a~priori$ with an
accuracy $\D x_i \sim l_{\hbox{\scriptsize Schwarzschild}}$. Now, if one measure one of
the spatial coordinate say
$x_1$, but the other coordinate $x_2$ was kept intact
such as $\D x_2 \sim l_{\hbox{\scriptsize Schwarzschild}}$, then the uncertainty relation (\ref{dxdx}) shows that the uncertainty
on the measured coordinate $x_1$ becomes
\beq
\label{dx=lf}
\D x_1 \geq \f{\theta}{2~\D x_2} \sim l_{\hbox{\scriptsize Compton}},
\eeq
where
\beq
\label{lf}
l_{\hbox{\scriptsize Compton}} =\f{\hbar}{2 mc}
\eeq
is (half of) the Compton length of the elementary black hole, i.e.,
one can not measure the coordinate of the elementary black hole with an
uncertainty smaller than $l_{\hbox{\scriptsize Compton}}$. 

This limit on the localization of the spatial
coordinate is the same as it for the elementary particle which does not
have substructure \cite{Land}. Furthermore, it seems that there is no
way to distinguish those two objects fundamentally: From the mass condition
(\ref{m<mc}) one obtains another condition,
\beq
\label{lf>lp}
l_{\hbox{\scriptsize Compton}} \geq  l_{\hbox{\scriptsize Planck}} 
\eeq
such as one obtains eventually 
\beq
\label{lf>lp}
l_{\hbox{\scriptsize Compton}} \geq  l_{\hbox{\scriptsize Planck}} 
\geq  l_{\hbox{\scriptsize Schwarzschild}}.   
\eeq
But according to (\ref{dx=lf}), this 
inequality means that one can
not localize the elementary black hole as much as one can measure its boundary  ($l_{\hbox{\scriptsize Compton}} \geq l_{\hbox{\scriptsize Schwarzschild}}$), which is consistent with the concept of the
elementary particle, and non-commutative space ($l_{\hbox{\scriptsize Compton}} \geq  l_{\hbox{\scriptsize Planck}}$) such as {\it the
  non-commutative space is not measured directly}.

A remarkable point of my derivation
is that I have not used any of the commutation relations for other dynamical
variables, i.e., $[x_i, p_j],~[p_i,p_j]$ $explictly$, which is contrast
to the usual derivation in the quantum mechanics \cite{Land} where the
commutation relation $[x_i,p_j]=i\hbar \delta_{ij}$ and $c$ as a limit
velocity\footnote{If one allow the ``tachyon'', which does not have
  a complete interpretation yet though, this
  quantum mechanical
  derivation will
  be invalid \cite{Park:96}.}are used. This fact
implies that one may consider generalized commutation relations
such as\footnote{Operator forms of $f_{ij},~g_{ij}$ will be strongly
  constrained by the Jacobi identities (see for example
  \cite{Synd}). But, the status of the Jacobi identities on the 
  non-commutative plane remains unclear.}   
\beq
\label{xp}
&&[x_i,p_j]=f_{ij} (\hbar, \theta),\\
\label{pp}
&&[p_i,p_j]=g_{ij} (\theta)~~(i=1,2).
\eeq
The commutation relation (\ref{xp}) shall be the usual one $[x_i,p_j]=i
\hbar$ when one use the current value of $G$ as $6.67 \times
10^{-8}~ \mbox{cm}^3 \mbox{g}^{-1} \mbox{sec}^{-2}$ for consistency with the quantum mechanical
derivation for the elementary particles such as the existence of an elementary black hole
is not inconsistent with the existence of an elementary particle. But since my derivation is independent on the value of $G$,
the additional contribution of $G$ through $\sqrt{\theta}=l_{\hbox{\scriptsize Planck}}$ need not
be always neglected: This may be important in the early universe with
the $time-varying-G$ scenario; furthermore, in this scenario the
non-commutative parameter change as $\theta (t) \sim G(t)$ if $\hbar,
c$ are the true universal constants.  

Moreover, since the critical mass $m_{NC}$ of (\ref{mc}) is large enough
such as all known elementary particles can be included in this category, the derivation of (\ref{dx=lf}) from the quantum theory
around the black hole on the non-commutative space, which should be
also understood as a quantum gravity, is sensible also for the elementary
particles if we identify the elementary black hole with an elementary
particle. In other words, the hierarchy of the length scales between $l_{\hbox{\scriptsize Compton}}$ and
$l_{\hbox{\scriptsize Planck}}$ is understood by considering the
quantum gravity on non-commutative space for
the elementary black hole (see Ref. \cite{Rand} for other mechanism
from small extra dimension); in particular, for
the case of $W^{\pm}, Z$ bosons which satisfy the mass condition (\ref{m<mc}) [$m_{W,Z}
\sim 10^{-22}~\mbox{g}$] well, the hierarchy is 
\beq
l_{\hbox{\scriptsize Compton}}&\sim& 10^{-16}~\mbox{cm}, \no\\
l_{\hbox{\scriptsize Planck}} &\sim& 10^{-33}~\mbox{cm} , \no\\
l_{\hbox{\scriptsize Schwarzschild}}&\sim& 10^{-50}~ \mbox{cm} \no
\eeq 
and this is explained essentially as 
\beq
l_{\hbox{\scriptsize Compton}} \times l_{\hbox{\scriptsize Schwarzschild}} \sim l_{\hbox{\scriptsize Planck}}^2.
\eeq
In other words, {\it the hierachy is a result of the space/space non-commutavity at the Planck scale and the
existence of the black hole horizon for the
elementary particle.}  
Moreover, if one consider the time-varying $\theta$ according to
the time-varying $G$ scenario, {\it there may be no hierarchy in the
early universe} (with $G \sim 10^{26}~ \mbox{cm}^3 \mbox{g}^{-1}
\mbox{sec}^{-2}$).  

All the results that I have considered so far supports the identification of the elementary black hole with an
elementary particle. Now, the questions, which might cause this
identification to be inaccurate, are\\

1) Can't the elementary black hole be probed below $l_{\hbox{\scriptsize
   Compton}}$ by any successive measuremnets contrast to an elementary
   particle ?\\

2) Is there the Hawking radiation which might cause the elementary
   black hole to be unstable contrast to the stable elementary particle ?\\

In order to consider the first question, one {\it needs} also to consider the space/time
non-commutativity\footnote{This has been considered in the context of
  string theory \cite{Yone} such as the narural non-commutativity scale
  $\sqrt{\theta}$ is the string length $l_s\sim \sqrt{ 2\pi
    \alpha'}$. But, in our context, this is considered
  as a generic property of particle theory.} 
\beq
\label{tx}
[t,x_i]=i \theta/c
\eeq 
in conformity with the quantum mechanics \cite{Land} on the limitation of
the time measurement:
\beq
\label{dt}
\D t \geq \f{\theta}{2~\D x_i} \sim l_{\hbox{\scriptsize Compton}}/c~
\eeq
[the elementary black hole is assumed also to have a best localised
position such as $\D x_i \sim l_{\hbox{\scriptsize Compton}}$ a priori before
the time measurement as in the measument of coordinate];
in this case, the 
causality may not be applied below the Compton length $l_{\hbox{\scriptsize Compton}}$ 
where the concept of time is fuzzy by the amount of $\D t \sim l_{\hbox{\scriptsize Compton}}/c$, the
time it takes light to traverse the distance $l_{\hbox{\scriptsize Compton}}$
\cite{Seib:00}. This explains why the two orthogonal coordinates $x_1,~x_2$ were not 
$simultaneously$ measured in (\ref{dx=lf}). Moreover, since the
operator nature of the time $t$ is a genuine fact in my framework of
ideas, there is no obstacle to understand the usual
energy-time uncertainty relation 
\beq
\label{dEdt}
\D E \D t \geq \hbar/2
\eeq 
as an operator relatrion anymore, which has been
absent in the conventional quantum mechanics,
\beq
\label{tE}
[t, E]=i \hbar
\eeq 
that is consistent with the non-commutative space-times commutation
relations (\ref{xx}) and (\ref{tx}). 

Now, let me show that how this energy-time uncentainty relation insure that $\D
x_i \geq l_{\hbox{\scriptsize Compton}}$ for any successive measurments in time $\D t$
after the first measurment of one coordinate $x_1$: From the
energy-time uncertainty relation (\ref{dEdt}) and limit on $\D t$
(\ref{dt}) one obtains $\D E=v \D p \sim mc^2$ such as $\D p
\sim mc,~ v \sim c$, where $\D p, v$ are the momentum uncertainty,
velocity of the particle after the first measurement in respectively; then, one finds
that the struck particle will move $\D x_i \sim v \D t \geq
l_{\hbox{\scriptsize Compton}}$ from the first observation point. $\D x_1$ is the same
order of the first measurement and so it is a repeatable measurment. But
the important point is that the other coordinate $x_2$, which has not been
measured in the first measuremnt but only considered a priori as the horizon size $\sim
l_{\hbox{\scriptsize Schwarzschild}} (< l_{\hbox{\scriptsize
    Compton}})$ which would be a genuine size of the
particle, can not be probed below $l_{\hbox{\scriptsize Compton}}$
still in the successive measurements. This insures that
$l_{\hbox{\scriptsize Compton}}$ is a genuine bound on the space
localization. But notice that in this measureing process, the minimum
uncertainty state (\ref{dxdx}) is destroyed and a higher uncertainty state is
observed in the second measuremnet, i.e., $\D x_1 \D x_2 \geq
l^2_{\hbox{\scriptsize Compton}}/2 > l^2_{\hbox{\scriptsize Planck}}$. 

Now, let me turn to the question on the Hawking radiation which might
cause the elementary black hole to be unstable. It is well known that the
large mass black hole has the Hawking radiation according to the
semi-classical quantization which neglect the back reaction
\cite{Hawk:74}. But, its validity is questionable for this microscopic
black hole where the back reaction can not be neglected anymore. Moreover, the space outside the horizon is
non-commutative space such as the usual physical laws might be broken
down near the horizon\footnote{With a naive use of the elementary particles'
  mass, charge, and interpretation of $j$ as the spin $\sim \hbar$,
  horizon becomes {\it naked}. But, this
  might not be the case from this non-commutativity.}. On the other
hand, similar to the atomic physics, quantum
mechanics would be also crucial to make the elementary black hole be
stable against the radiation\footnote{This has been recently argued
  also by Adler {\it et al.} \cite{Adle} within the generalized
  uncertainty principle context.} ; but complete understanding needs to be
discovered\footnote{Probably, this elementary black hole will be
  relevant to the ground state of the macroscopic black holes \cite{Park:01}.}.

{\it Large Black Hole}:
For large black hole whose horizon radius $r_+$ is bigger than the
Planck length $l_{\hbox{\scriptsize Planck}}$ one can probe its
boundary $(l_{\hbox{\scriptsize Schwarzschild}}>l_{\hbox{\scriptsize
    Compton}})$ such as it can not be considered as an elementary
particle anymore. Moreover, the non-commutative space have neglizable
relavance in the dynamics for center of black hole $(l_{\hbox{\scriptsize Schwarzschild}}>l_{\hbox{\scriptsize Planck}})$ but rather have
dominant relavance in the surface dyanamics on the horizon
$(l_{\hbox{\scriptsize Planck}}>l_{\hbox{\scriptsize Compton}})$. This
may be understood as a holography of the bulk non-commutative spaces
onto the horizon of the semiclassical black hole \cite{Beke:94,tHoo:93,Suss:94}. 
So, it would be more natural to consider the compactified
noncommutative space for a treatment of the large black hole. 

As the
simpliest space, let me consider a non-commutative sphere with three coordinates
$x_i~ (i=1,2,3)$
\beq
[x_i, x_j]=i \frac{\theta}{r} \epsilon_{ijk} x_k,
\eeq
where $r$ is the radius of the non-commutative sphere. Now then, let me identify this non-commutative sphere as the black hole
horizon, then the discrete spectrum of the
horizon area $A$ is derived, with an identification of $x_i^2=r^2$, within the
non-commutative quantum mechanics \cite{Nair:01b}:
\beq
A&=&\oint_{r=r_+} d^2 x = 4 \pi r^2 \nonumber \\
 &=&4 \pi l_{\hbox{\scriptsize Planck}}^2 ~\sqrt{j (j+1)}, 
\eeq
where $j$ is the heightest weight of an irreducible representation of
$SU(2)$ spin algebra of $R_i=(r/\theta) x_i$. 

On the other hand, this
non-commutative sphere approach is exactly the same as what Bekenstein \cite{Beke:01}
has assumed in an algebraic formulation of the black hole quantization by
identifying $R_i$ and $A$ with Bekenstein's angular momentum operator
$\hat{\bf J}$ and area operator $\hat{A}$, in respectively. So, it will be an
interesting problem to derive the uniformly spaced area spectrum, $A\sim
l^2_{\hbox{\scriptsize Planck}} n$ \cite{Beke:74}, which produces a more
interesting interpretation on the  black hole entropy, within my context also without assuming
other more complicated algebraic works.


Finally, I note that, for the quantization of an elementary black hole
system, study on the
bulk non-commutative space is needed in contrast to large black
hole. In this procedure, finding the black hole solutions on the
non-commutative space will be a outstanding challenge which will be a first step toward the quantization \cite{Ho}.
\begin{center}
 {\bf Acknowledgment}
\end{center}

I would like to thank Parameswaran Nair and Alexios Polychronakos for comments. This work was supported in 
part by a CUNY Collaborative Incentive Research Grant.


\begin{thebibliography}{99}

\bibitem{Conn} There are a huge number of papers on diverse aspects. For an overall view of this field see A. Connes,
  M. R. Douglas and A. Schwartz, 
JHEP 9802, 003 (1998); 
N. Seiberg and E. Witten, 
JHEP 9909, 032 (1999). 

\bibitem{Dunn} 
S. Girvin and T. Jach,
  Phys. Rev. {\bf D29}, 5617 (1984); G. Dunne, R. Jackiw and C. Trugenberger,
  Phys. Rev. {\bf D41}, 661 (1990).

\bibitem{Suss:01} L. Susskind, hep-th/0101029; A. P. Polycronakos,
  hep-th/0103013; 
  D. Karabali and B. Sakita, hep-th/0106016.

\bibitem{Nair:01} 
M. M. Sheikh-Jabbari, Phys. Lett. {\bf B510}, 247 (2001); 
V. P. Nair and A. P. Polychronakos, Phys. Rev. Lett. {\bf 87} 030403
 (2001);
D. Bak, K. Lee and J.-H. Park, Phys. Rev. Lett. {\bf 87}, 030402
(2001). 

\bibitem{Beke:73} J. D. Bekenstein, Phys. Rev. {\bf D7}, 2333 (1973).

\bibitem{tHoo:96} G. 't Hooft, 
Class. Quant. Grav. {\bf 13}, 1023 (1996); 
Int. J. Mod. Phys. {\bf A11}, 4623 (1996). 


\bibitem{Land} L. D. Landau and R. Peierls, Z. Phys. {\bf 69}, 56
 (1931); V. B. Berestetskii, E. M. Lifschitz and L. P. Pitaevskii,
 {\it Quantum Electrodyanamics} (Pergamon Press, New York, 1982);
 K. Gottfried, {\it Quantum Mechanics, Vol.I} (W. A. Benjamin Inc.,New
 York, 1966).

\bibitem{Park:96} M.-I. Park and Y.-J. Park, Nuov. Cim. {\bf 111B},
  1333 (1996).

\bibitem{Synd} H. S. Snyder, Phys. Rev. {\bf 71}, 38 (1947);
  M. Maggiore, Phys. Lett. {\bf B319}, 83 (1993); S. de Haro, JHEP
  {\bf 10}, 023 (1998); N. Sasakura, JHEP
  {\bf 05}, 015 (2000).

\bibitem{Rand} 
L. Randall and R. Sundrum, 
Phys. Rev. Lett. {\bf 83}, 3370 (1999). 

\bibitem{Yone} T. Yoneya, 
hep-th/0010172.

\bibitem{Seib:00} N. Seiberg, L. Susskind and N. Toumbas, 
hep-th/0005015.

\bibitem{Adle} R. J. Adler, P. Chen, and D. I. Santiago, gr-qc/0106080. 

\bibitem{Park:01} M.-I. Park, 
hep-th/0111224.

\bibitem{Hawk:74} S. W. Hawking, Nature {\bf 248}, 30 (1974).

\bibitem{Beke:94} J. D. Bekenstein, 
Phys. Rev. {\bf D49}, 1912 (1994). 


\bibitem{tHoo:93} G. 't Hooft, 
gr-qc/9310026.

\bibitem{Suss:94} L. Susskind, 
J. Math. Phys. {\bf 36}, 6377 (1995). 

\bibitem{Nair:01b} 
D. Karabali, V. P. Nair and A. P. Polychronakos, hep-th/0011249.

\bibitem{Beke:01} J. D. Bekenstein, hep-th/0107045.

\bibitem{Beke:74} J. D. Bekenstein, Lett. Nuovo. Cim. {\bf 11}, 467
  (1974); V. Mukhanov, JETP Lett. {\bf 44}, 63 (1986); Y. I. Kogan,
  {\it ibid.}, {\bf 44}, 267 (1986); A. Alekseev, A. P. Polychronakos, and
  M. Smedb\"{a}ck, hep-th/0004036.


\bibitem{Ho} 
M. Ba\~nados, {\it et al.},
hep-th/0104264; V. P. Nair, hep-th/0112114.

\end{thebibliography}
\end{document}